\documentclass[runningheads]{llncs}
\usepackage{graphicx}
\usepackage{epstopdf}
\usepackage[misc]{ifsym}
\usepackage{amsfonts}
\usepackage{amsmath}
\usepackage{multirow}
\usepackage{hyperref}
\usepackage{pdfpages}
\usepackage{cite}
\hypersetup{hypertex=true,colorlinks=true,linkcolor=blue,anchorcolor=blue,urlcolor=blue,citecolor=blue}

\begin{document}
\title{Multiscale Autoencoder with Structural-Functional Attention Network for Alzheimer's Disease Prediction}
%
%

\author{Yongcheng Zong \and
Changhong Jing \and
Qiankun Zuo \textsuperscript{(\Letter)}}

\authorrunning{Y. Zong et al.}

\institute{Shenzhen Institutes of Advanced Technology, Chinese Academy of Sciences, Shenzhen
	518000, China\\
	\email{\{yc.zong,ch.jing,qk.zuo\}@siat.ac.cn}
}

%
%
%

%

%
%
%
\maketitle              
\begin{abstract}
The application of machine learning algorithms to the diagnosis and analysis of Alzheimer's disease (AD) from multimodal neuroimaging data is a current research hotspot. It remains a formidable challenge to learn brain region information and discover disease mechanisms from various magnetic resonance images (MRI). In this paper, we propose a simple but highly efficient end-to-end model, a multiscale autoencoder with structural-functional attention network (MASAN) to extract disease-related representations using T1-weighted Imaging (T1WI) and functional MRI (fMRI). Based on the attention mechanism, our model effectively learns the fused features of brain structure and function and finally is trained for the classification of Alzheimer's disease. Compared with the fully convolutional network, the proposed method has further improvement in both accuracy and precision, leading by 3\% to 5\%. By visualizing the extracted embedding, the empirical results show that there are higher weights on putative AD-related brain regions (such as the hippocampus, amygdala, etc.), and these regions are much more informative in anatomical studies. Conversely, the cerebellum, parietal lobe, thalamus, brain stem, and ventral diencephalon have little predictive contribution.
\end{abstract}

\section{Introduction}
Alzheimer's disease (AD) is a very serious and irreversible neurodegenerative disease among old people\cite{2009Reviews}. The diagnosis of Alzheimer's disease is a difficult task for neurologists, and traditional test scales such as memory tests cannot accurately determine the disease type \cite{2014Biomarker}. With the help of machine learning algorithms emerging in recent years, computer-aided diagnosis has been widely used in the field of medical image analysis and achieved great sucess\cite{wang2018bone,hu2019cross}. One of the major challenges in developing brain image-based AD diagnostic methods is the difficulty in visually seeing AD-induced changes in the brain due to the very complex structure of brain tissue.

Magnetic resonance imaging is a non-invasive, repeatable, and high spatial resolution technique that is widely used in the diagnosis and research of various brain diseases \cite{wang2020ensemble}. Among them, fMRI imaging based on the BOLD effect, which uses the blood oxygen level-dependent signal as a neurophysiological indicator, is widely used in the identification of neurodegenerative diseases \cite{2017Resting}, especially in the classification of MCI and AD stages. The T1-weighted magnetic resonance imaging (T1WI) is used to extract brain tissue morphological features among the whole brain to capture volume change information in different brain regions. Since different modal images carry different disease-related information, researchers explore methods using T1WI or fMRI for disease diagnosis. For example, the work in \cite{yu2021morphological} developed an MRI-based model and achieve good performance in AD diagnosis; Xiao et al. \cite{ref_6} utilized fMRI data to fully explore functional features and gain high detection accuracy. As multimodal images can provide complementary information about the disease, recent studies \cite{yu2020multi,ref_7,ref_7_1} with multimodal data have shown better performance compared with methods using unimodal data. Therefore, we make use of the T1WI and fMRI in our model to conduct multimodal feature fusion for disease diagnosis.

In recent years, deep learning has achieved great success in medical image analysis \cite{ref_1,ref_2,ref_2_1,ref_2_3,ref_2_4,ref_2_5}, especially the famous convolution neural network (CNN)\cite{ref_3,ref_4_0,ref_4,ref_5,ref_5_1,ref_5_2}. It automatically learns the best features from data distribution because of its strong ability in local feature extraction. It was shown by Ju et al. \cite{Ronghui2017Early} that the autoencoder model achieved better performance in diagnosis compared to methods such as support vector machines, logistic regression, linear discriminant analysis, etc. Most CNN-based segmentation models have an encoder-decoder structure, such as the fully convolutional network (FCN) \cite{2015Fully}. The FCN first extracts feature through a series of convolutional layers and restore the size of the original image using the convolutional layers as decoders. This end-to-end mechanism enables the network to generate full-scale segmentation outputs. Meanwhile, the Generative adversarial network (GAN)\cite{2014NEURAL} has become the most popular deep generation model in the fields of medical image recognition \cite{ref_article09} since it can generate realistic data without explicitly modeling the data distribution. GAN can bee seen as variational-inference \cite{ref_article10a1,ref_article10a2,ref_article10a3,ref_article10a4} based generative model and has also been widely applied in the Alzheimer's disease diagnosis, such as image generation \cite{hu2020brain,hu2021bidirectional}, image recognition \cite{wang2020patent,yu2021tensorizing}, image segmentation \cite{ref_10}, image super-resolution \cite{ref_11}. Network analysis has been employed widely in kinds of areas \cite{wang2015hadoop, wang2012random, wang2011quantitative, w2012defining} and brain network analysis is a potential tool for brain science. GAN is also used for brain network analysis.
\cite{pan2021decgan,zuo2021pggan,pan2021characterization}. The GANs can extract latent representations from brain imaging for AD recognition and prediction. However, different fusion strategies for the processing of the global and local features can affect the identification of key brain regions in AD \cite{pan2021characterization}, and influence the model's classification performance.

This paper proposes an end-to-end multiscale autoencoder-based Alzheimer's disease prediction network, which consists of three modules, a feature extraction network, a feature fusion network, and a classifier. In the feature extraction stage, the T1WI data is processed in patches and then input to the corresponding autoencoder modules, the two patch-level modules (sPM, fPM) for T1WI and fMRI, respectively. We improve Alzheimer's disease detection performance by leveraging learned multimodal complementary features and intramodal correlation information. The multimodal fusion network of our proposed self-attention mechanism plays a key role, where channel attention aims at dimensionality reduction and spatial attention aims at identifying the most important information. The mechanism effectively fuses the multimodal data, thereby improving the classification performance. Experiments on the Alzheimer's Disease Neuroimaging Initiative (ADNI) database show that compared with the H-FCN without the autoencoder structure, our method has achieved better results in AD classification.

The main innovations of this paper are as follows:

(1) Our model successfully learns the heterogeneity of different individual brain networks. The automatic alignment and brain atlas registration of general preprocessing software cannot adapt to the different brain structures of a large group of people. The preprocessed data is often based on the average value between regions, which loses a lot of potential information on the voxels.

(2) The multiscale autoencoder of the feature extraction network in our model can gain significant features by extracting features from different levels of sub-networks, which contribute to better classification prediction performance. We directly use 3D T1WI data and provide a visual representation of embedding, so the results have better interpretability.

(3) The self-attention mechanism integrates the data features of the two modalities in practice. Based on the visual display of the weights under different binary classification tasks, we obtain areas of structural changes that are consistent with the clinical practice, which is beneficial for disease diagnosis.

\begin{figure}
\includegraphics[width=\textwidth]{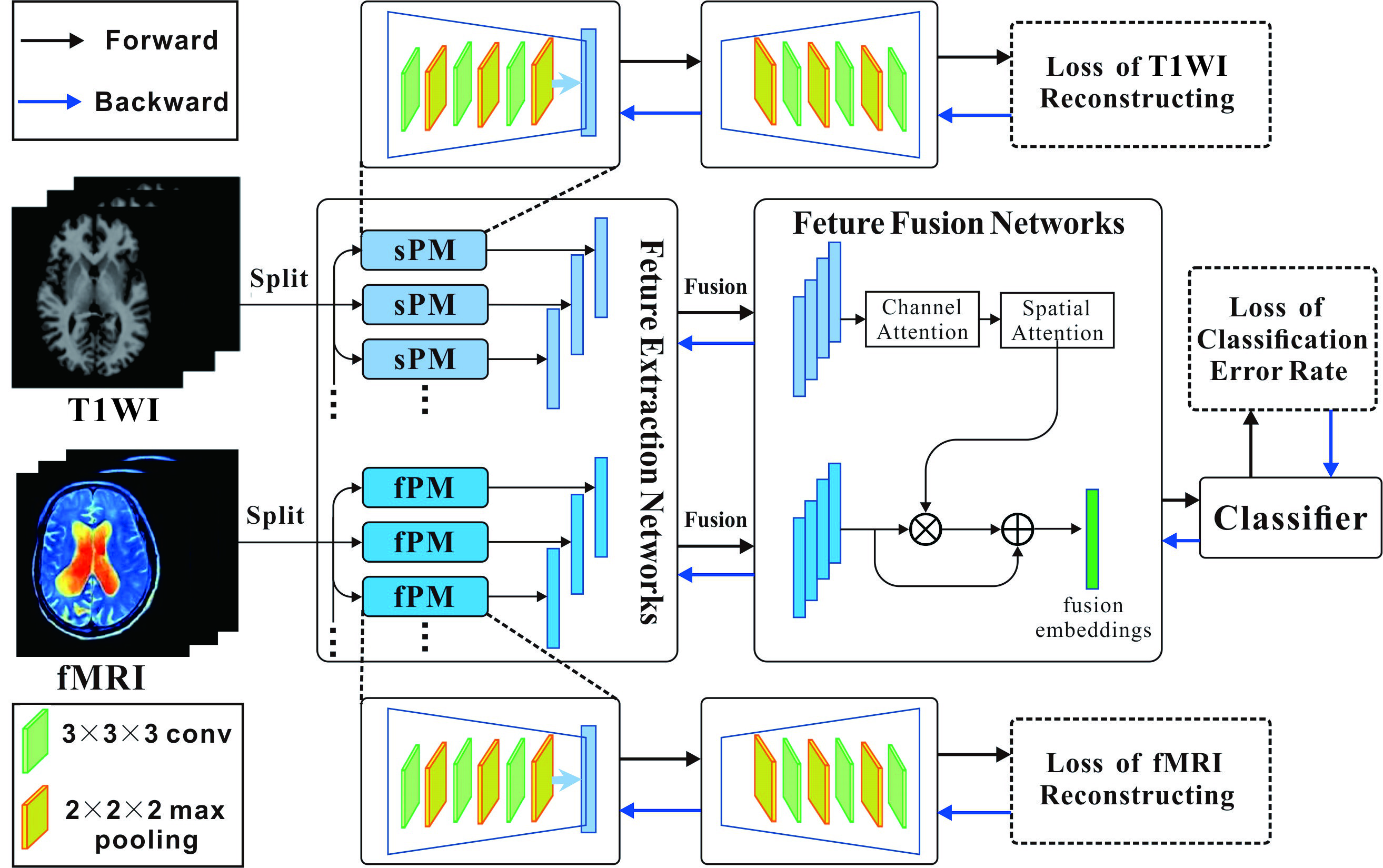}
\caption{The architecture of the proposed model.} \label{fig1}
\end{figure}


\section{Method}
Fig.~\ref{fig1} shows the overall framework of the model proposed in this paper. The model consists of three main modules, namely, the multiscale autoencoder, the multimodal feature fusion with attention mechanism, and the final classification MLP network. In the feature extraction stage, the T1WI data is processed in patches and input to the corresponding autoencoder modules, that is, patch-level modules (sPM, fPM) for T1WI and fMRI, respectively. The network parameters of each module are independent of each other, so each module shares the same values of weights. Based on the learned features and fused features, it is finally used for the classification of AD subjects.

\subsection{Feature Extraction Network: Autoencoder} 
Inspired by the successful application of Transformer \cite{2020An} in CV, this paper also processes T1WI raw data into patches in the beginning. Different from \cite{zhao2021region} to divide the brain according to putative brain anatomy,  we directly divide the brain image into $4\times4\times4$ non-overlapping sub-regions in three dimensions, and each patch is processed by a separate sub-network to extract the region features of the corresponding part. By using convolution downsampling operations three times, the sPM reduces the original dimension of the image to the size of $2\times2\times2$ in the first stage. Based on the analysis, the intermediate embedding results can be easily visualized.

The loss function of this stage is the reconstruction loss between the original and generated T1WI. To extract sparse features \cite{2021Temporal}, a sparse penalty term for encoding is added to the loss function:

\begin{equation}
\Omega(h)=\lambda \sum_{i}\left|h_{i}\right|
\end{equation}
where h is the encoding parameter.

The structure of the fPM and sPM design is similar. For fMRI data, the extra dimension of the time is added to the input. At last, we get features of each region of the brain. These features are then fused by the self-attention mechanism we propose.

Guided by image reconstruction loss \cite{2021A}, we first pre-train the feature extraction network of our model.

\begin{equation}
\left\{\begin{array}{l}
L_{s}=\frac{1}{N} \sum_{i}^{N}\left\|\widehat{A s}_{i}-A s_{i}\right\|_{2}^{2}+\Omega(h) \\
L_{f}=\frac{1}{N} \sum_{i}^{N}\left\|{\widehat{A f_{i}}}-A f_{i}\right\|_{2}^{2}+\Omega(h)
\end{array}\right.
\end{equation}
where $\widehat{A s}_{i}$ i  and $\widehat{A s_{i}}$ represent the generated and original T1 structure imaging, respectively and $\widehat{A f}_{i}$ i  and $\widehat{A f_{i}}$ i represent the generated and original fMRI imaging, respectively.

\subsection{Feature Fusion Network} 
Self-attention fusion is carried out in two dimensions of spaces and channels, and the self-attention calculation of each region is processed as follows:

\begin{equation}
\left\{\begin{array}{c}
\text { key }_{i}=\text { region }_{i} \cdot W^{(k)} \\
\text { value }_{i}=\text { region }_{i} \cdot W^{(v)} \\
\text { query }_{i}=\text { region }_{i} \cdot W^{(q)}
\end{array}\right.
\end{equation}
where $ \text { region }_{i}$ represents the initial output of $i$th patch of brain imaging, $W^k$, $W^v$, and $W^q$ are convolution layers.

Update the output through the self-attention mechanism:
\begin{equation}
\text { weight }_{ij}=softmax( \text{query}_{i} \cdot \text {key}_{j})
\end{equation}
\begin{equation}
\text { new\_region }_{i}=\sum_{j}^{r}\left(\text { weight }_{i j} \cdot \text { value }_{j}\right)
\end{equation}

To effectively utilize both T1-weighted structural features and fMRI functional features, we must fuse these two different modalities \cite{zuo2021multimodal}. In addition, any redundancy and information that is not important to either mode must be reduced. We design a multimodal module with an attention mechanism to improve the compatibility of two different modalities and then distinguish fusion features by a classifier for subsequent classification tasks.

\begin{figure}
\includegraphics[width=\textwidth]{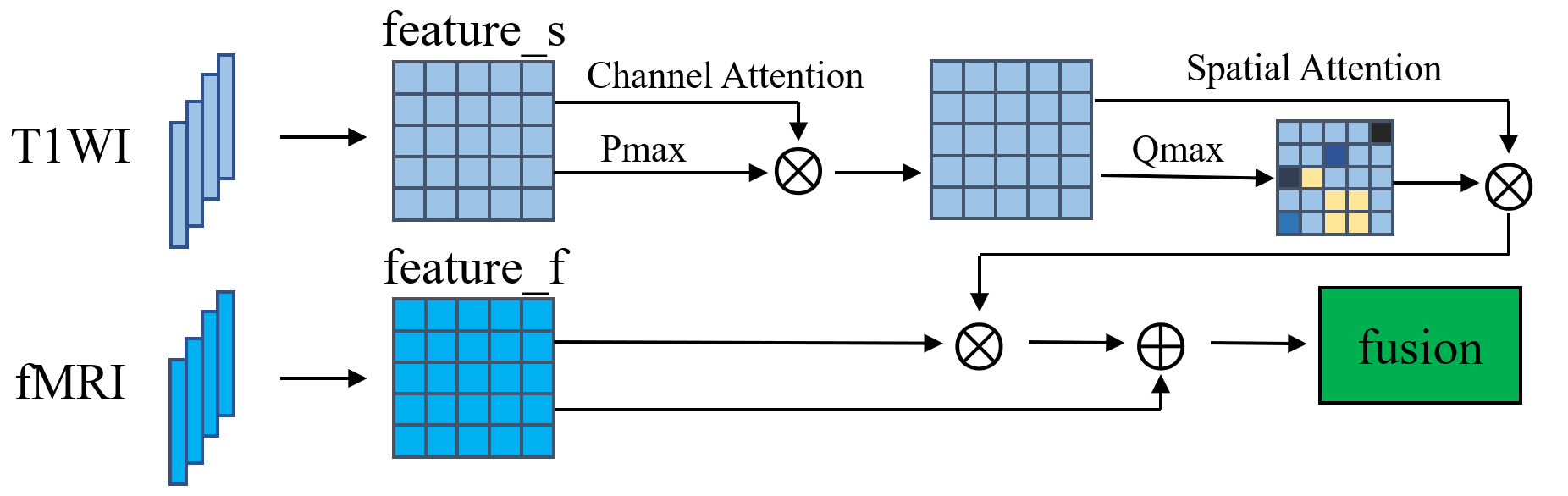} 
\caption{The proposed feature fusion mechanism. Pmax represents the global max-pooling operation and Qmax denotes the global max pooling along the space axis. The T1 features are passed through Channel Attention and Spatial Attention to locate what and where is significant.} \label{fig2}
\end{figure}

Learned from \cite{2021Deep}, we improve the original channel attention and spatial attention method. The fusion process of extracted features can be expressed by the following formula:

\begin{equation}
\left\{\begin{array}{l}
\boldsymbol{f}_{\mathrm{T} 1}^{c}=\text { Attention }{ }^{c}\left(\boldsymbol{f}_{\mathrm{T} 1}\right) \in \mathbb{R}^{H \times W \times C} \\
\boldsymbol{f}_{\mathrm{T} 1}^{S}=\text { Attention }^{s}\left(\boldsymbol{f}_{\mathrm{T} 1}^{c}\right) \in \mathbb{R}^{H \times W \times C} \\
\boldsymbol{f}_{f u s i o n}=\boldsymbol{f}_{f M R I} \otimes \boldsymbol{f}_{\mathrm{T} 1}^{S} \oplus \boldsymbol{f}_{f M R I}
\end{array}\right.
\end{equation}
where $\oplus$ denotes element-wise addition, and $\oplus$ denotes element-wise multiplication. By combining T1 weighted features, the fusion feature will contain extra complementary information. We use a T1-guided attention module at each level of the feature fusion network to obtain the aggregation features. The specific mechanism of the feature fusion process is illustrated in Fig.~\ref{fig2}.

\subsection{Classifier Network}	
Considering that the full connection network has good classification capabilities and generalization, we choose the MLP network with easy implementation and high efficiency to calculate the classification regression loss with the labels of subjects.

We use cross-entropy as the loss function of the classifier. To be more specific, the loss function to this end is the combination of the loss for weakly-supervised discriminative localization and the loss for multi-task regression \cite{2019End}

\begin{equation}
\mathrm{L}_{reg}=-\frac{1}{N} \sum_{n=1}^{N} \sum_{c=1}^{C} \mathbf{1}\left(\mathrm{y}_{n}=c\right) \log \left(\mathrm{y}_{n}\right)
\end{equation}
where $\mathbf{1(\cdot)}$ is a binary indicator and $y_n$ is the ground truth.

The total loss function is as follows:
\begin{equation}
\mathrm{L}=\alpha \mathrm{L}_{s}+\beta \mathrm{L}_{f}+\mathrm{L}_{reg}
\end{equation}
where $\alpha$, $\beta$ are the super parameters, and both are set to 0.5.

\section{Experiments}
\subsection{Dataset}
The experiments are performed on the public Alzheimer's Disease Neuroimaging Initiative (ADNI) dataset \cite{2010ADNI}. We use the ADNI3 dataset and download 206 pairs of T1WI and fMRI imaging datasets in total. T1WI is a 3D MR image with a voxel resolution of 240$\times$256$\times$176. fMRI is a 4D MR image, the original format is 704$\times$704$\times$976 time slices, after conversion and removal of boundary padding, the input resolution is converted to 976$\times$60$\times$64$\times$44. The dataset contains 72 AD, 34 EMCI, 36 LMCI, and 64 NC samples, respectively. 70\% of the data is used as the training set, and the remaining 30\% is used to test the classification performance. Since the same weight is assigned to the structural and functional brain networks,  there is no validation set to adjust the hyperparameters of the model.

\subsection{Experimental Settings}
The model is trained on the TensorFlow1 platform. The data is directly fed into the network for training without general preprocessing methods such as skull stripping or registration. The NVIDIA TITAN RTX2080 GPU device is used to train the model for 24 hours with about 800 epochs. The learning rate is set to 0.001 initially and the network parameters are updated by Adam optimizer \cite{2014Adam}. Each layer of the network adopts group normalization and ReLU activation function in the feature extraction network. The classifier adopts a fully connected layer, and finally performs classification prediction through a softmax layer. The $\alpha$ and $\beta$ in our experiments is empirically set to 0.5 and 0.5, respectively.

The sPM network parameters are listed in the following Table~\ref{tab1}, where GN denotes Group Normalization, and AddId denotes a residual structural connection. Like the encoder, the decoder has a symmetrical structure. The difference is that the downsampling of convolution with stride 2 in the encoder process is replaced by the upsampling of trilinear interpolation in the decoder. Other structures remain unchanged and are not listed in detail. It is worth noting that we also add concatenations between the encoder and decoders.

\begin{table}
\centering
\caption{The settings of the sPM network parameters.}\label{tab1}
\begin{tabular}{p{2.5cm}p{4.5cm}l}
\hline
Name &  Ops & Output size\\
\hline
Input, T1WI &   & 1$\times$240$\times$256$\times$176\\
InitConv & Conv & 32$\times$240$\times$256$\times$176\\
EncoderDown1 & Conv stride2 & 64$\times$120$\times$128$\times$88\\
EncoderBlock1 & [GN, ReLU, Conv]$\times$2, AddId & 64$\times$120$\times$128$\times$88\\
EncoderDown1 & Conv stride2 & 128$\times$64$\times$64$\times$44\\
EncoderBlock1 & [GN, ReLU, Conv]$\times$2, AddId & 128$\times$64$\times$64$\times$44\\
EncoderDown1 & Conv stride2 & 128$\times$30$\times$32$\times$22\\
EncoderBlock1 & [GN, ReLU, Conv]$\times$2, AddId & 64$\times$30$\times$32$\times$22\\
\hline
\end{tabular}
\end{table}

\subsection{Results} 
\subsubsection{Performance comparison} 

\begin{figure} 
\centering
\includegraphics[width=0.8\textwidth]{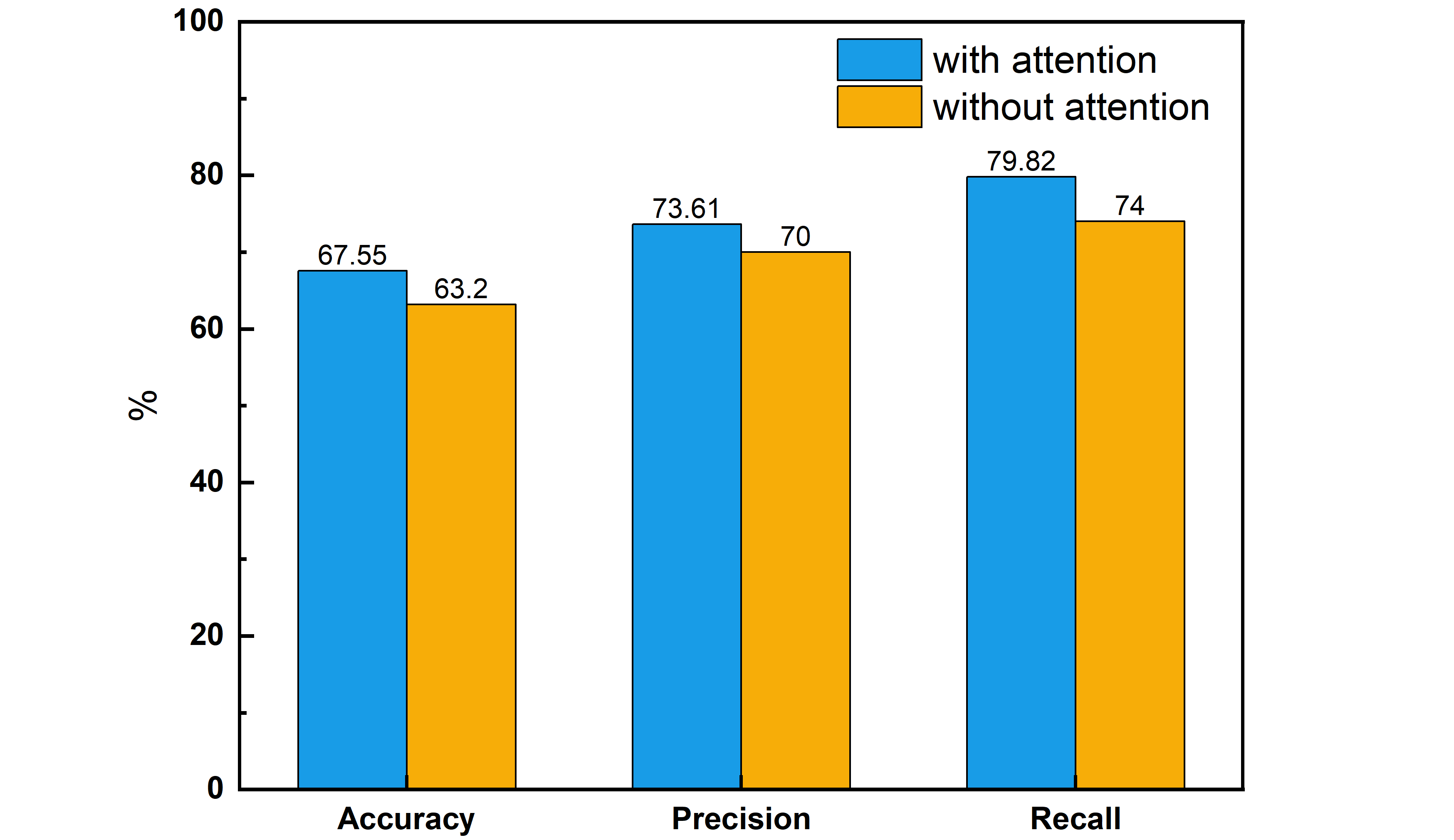} 
\caption{The comparison of attention mechanism(with/without). The left blue bins denote the performance of our model, and the right origin bins denote the other method by simply element-wise addition.} \label{fig3}
\end{figure}

The experiment performance is evaluated according to three metrics: classification accuracy, precision, and recall. We choose H-FCN as the baseline comparison, and CNN is used to hierarchically identify patch-level and region-level identification locations in whole-brain T1WI automatically since CNN can obtain good performance in brain tumor identification \cite{2015U}. The accuracy and the recall rate are calculated by treating AD as the positive class (AD vs. NC and MCI).

\begin{table}
\centering
\caption{The prediction performance in three metrics. The training/testing sets ratio is 7:3. Values are reported as mean $\pm$std deviation.}\label{tab2}
\begin{tabular}{p{1.5cm}p{2cm}p{2cm}p{2cm}}
\hline
Method & Accuracy & Precision & Recall\\
\hline
H-FCN & 64.23$\pm$5.33 & 68.35$\pm$2.54 & {\bfseries 82.55$\pm$4.73} \\
Ours & {\bfseries 67.55$\pm$4.26} & {\bfseries 73.61$\pm$3.16} & 79.82$\pm$4.49 \\
\hline 
\end{tabular}
\end{table}

Based on the well-trained model, we have conducted 10 experiments in total in the test dataset. The best and largest results are labeled in bold in each row. It can be seen from Table~\ref{tab2} that the variance difference between the two methods is not large, reflecting the stability of the two methods is basically the same. In general, except for recall, which lags behind H-FCN, all other performance metrics lead by 2-3 percentage points. The proposed model behaves in poor performance in the recall metric evaluation.

\begin{figure}
\includegraphics[width=\textwidth]{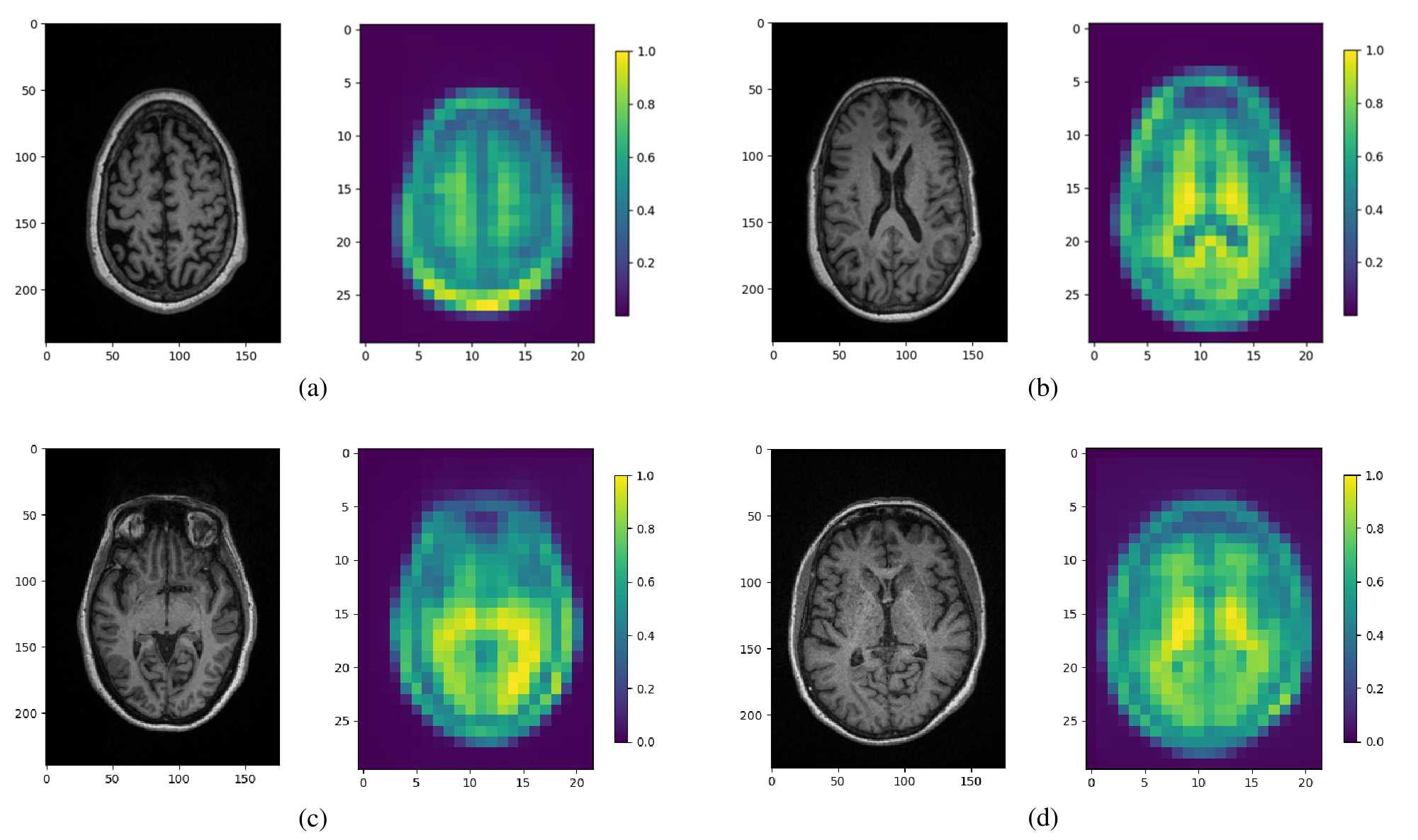} 
\caption{Embedding visualization. The left is a T1WI slice of one subject, and the right is the weight values of the fusion embedding.} \label{fig4}
\end{figure}
\subsubsection{Ablation Study} 

The input of the experiment includes brain image data of two modalities. To verify the effect of our fusion mechanism on classification accuracy, we designed a set of ablation experiments. The following chart in fig.~\ref{fig3} demonstrates the results by comparing with another simple strategy of element-wise addition on two features. All methods are trained using the same AD and NC subjects and also are tested on the same testing data. We can conclude that the proposed fusion mechanism does work by improving the classification performance in all three evaluation metrics.

\subsubsection{Feature visualization} 
Fig.~\ref{fig4} shows the visual results of the same subject's brain sections and the corresponding extracted embedding. Some edge portion of the embedding visualization has a higher value which should be omitted (since original imaging data has not been pre-processed to eliminate the scalp regions). The weight values in the area of the cerebral cortex are distributed randomly overall, and there is a highlighted part (yellow) in the hippocampus area associated with the disease.

By visualizing the extracted features, we find that the hypothesized AD-related brain regions (such as the hippocampus, amygdala, etc.) have higher weights, and these regions contain more information. In contrast, the cerebellum, parietal lobe, thalamus, brain stem, and ventral interbrain control little predictive effect.

\section{Conclusion}
In this paper, we propose an end-to-end framework for Alzheimer's disease diagnosis based on a multiscale autoencoder. Specifically, a CNN-based encoder is used to extract features in the brain image representation space, and then attention-based fusion modality features are utilized for AD detection and brain region identification. The results on the ADNI dataset show that our method can effectively improve the classification performance, and the regions with high thresholds reveal abnormal brain regions in AD patients. In the future, we will explore the intrinsic mechanisms of multimodal fusion and extend this work to multi-task classification.

%
%
%
%

\end{document}